**Short Paper**

# Internet Security Awareness of Filipinos: A Survey Paper

Challiz D. Omorog
College of Information and Communications Technology, Camarines Sur Polytechnic Colleges
Camarines Sur, Philippines
challizomorog@cspc.edu.ph
(corresponding author)

Ruji P. Medina
Graduate Studies, Technological Institute of the Philippines
Quezon City, Philippines
ruji_p_medina@yahoo.com

*Date received*: January 17, 2018
*Date received in revised form*: March 1, 2018
*Date accepted*: March 20, 2018

Recommended citation:

Omorog, C. D., & Medina, R. P. (2017). Internet security awareness of Filipinos: A survey paper. *International Journal of Computing Sciences Research*, 1(4), 14-26. doi: 10.25147/ijcsr.2017.001.1.18.

**Abstract**

*Purpose* – This paper examines the Internet security perception of Filipinos to establish a need and sense of urgency on the part of the government to create a culture of cybersecurity for every Filipino.

*Method* – A quantitative survey was conducted through traditional, online and phone interviews among 252 respondents using a two-page questionnaire that covers basic demographic information and two key elements – (1) Internet usage and (2) security practices.

*Results* – Based on findings, there is a sharp increase of Internet users for the last three years (50%) and most access the Internet through mobile (94.4%). Although at home is the most frequent location for Internet access (94.4%), a good percentage still use free WiFi access points available in malls (22.2%), restaurants (11.1%), and other public areas (38.9%) doing Internet services (email and downloading) that are vulnerable to cyber attacks. The study also revealed that although respondents may have good knowledge of Internet security software, proper implementation is very limited.

*Conclusion* – Filipinos are susceptible to cyber attacks, particularly to phishing and malware attacks. Also, majority of the respondents' Internet security perception is derivative- they practice online measure but with limited understanding of the purpose. Therefore proper education, through training and awareness, is an effective approach to remedy the situation.

*Recommendations* – The Philippine government must now take actions and tap industries to educate Filipinos about Internet security before any negative consequences happen in the future.

*Research Implications* – The information collected sets a clear picture on the importance of cybersecurity awareness from a regional to a global perspective.

## INTRODUCTION

According to the Internet World Stats (Miniwatts Marketing Group, 2017), and Hootsuite and We Are Social (Kemp, 2017), the Internet audience in the Philippines scaled up to 60 million as of January 2017. It also reported that the Philippines has the highest Internet penetration growth rates with no signs of deceleration. This growth has become the launchpad of the government to start-up and expand their operations online (e.g., www.gov.ph), influencing many sectors such as business, academe, and health.

E-commerce is also eyed as the major economic growth propeller in the country, which is estimated to gain more grounds (Department of Trade and Industry, 2016). Engagements that require credit cards usage, online banking transactions, and electronic data interchange are also expected to heighten in a significant increase. However, this sophistication and convenience also fueled a number of security threats in the Philippines. Cybercrime cases such as online scams, identity thefts, and cyber attacks on banks have occasionally been reported online and in local media news. For example, in August 2016, the Department of Health (DOH) website was hacked and used to phish out sensitive information from Bank of Philippine Islands (BPI) cardholders. Also in April 2016, the Bangko Sentral ng Pilipinas (BSP), the central bank of the Philippines, reported a cyber attack.  The BSP governor claims that the attack was limited to the BSP website only and not its banking system.

These reports point to how pervasive cyber threats have become. However, contrary to popular belief that hackers target only big companies, attacks have been increasingly common to home and small office (HOSO) routers as well. This is because hackers hunt poorly managed computer systems (Pan, Zhong, & Mei, 2015) with varied intentions frequently for monetary gain (Australian Institute of Criminology, 2005). So end-users, with the least effort and awareness on cybersecurity landscape, are considered the "weakest link" and the frequent target of cyber attacks (Aloul, 2012). The lack of IT security know-how and skillset among Filipino end-users will incessantly make the Philippines prone to cybersecurity threats (Microsoft, 2017).

The primary purpose of this paper is to provide empirical evidence and consciousness to the Philippine government on the need for cybersecurity education, training and awareness programs, as well as motivate the country to take action to create a culture of cybersecurity in every "*Juan de la Cruz*". The information collected sets a clear picture on the importance of cybersecurity from a global to a regional perspective through the existence of legal policies, capability-building, organizational and collaborative cooperation.

## LITERATURE REVIEW

### *Security Measures in Wireless Networks*

The established popularity of wireless networks among Internet users particularly in HOSO environment is a testament to its greater flexibility and integration prowess. Access to the latest wireless technology was made possible because of the inexpensive wireless access points (WAPs) in the market pre-equipped with the standard network security protocols. In the Philippines alone, pocket WAPs are sold in regular stores for less than Php 1,000 (US$ 20) only, intended for students, and people on the go. Users get immediate Internet connection, relying entirely on the WAPs or Internet service providers' (ISPs) default security systems. Theydo not bother to read the access point manual to change the default administrator credentials of the router as long as a connection is established. The carelessness of users (Hasan, Rahman, Abdillah, & Omar, 2015) and the open nature of the wireless network provide opportunities for cyber attackers (Zou, Zhu, Wang, & Hanzo, 2016).

Currently, Wi-Fi Protected Access 2 (WPA2) protocol, from its predecessor WPA, offers the standard encryption mechanism for WAPs (Scarpati, 2009). The papers of Sari and Karay (2015), and Ijeh, Brimicombe, Preston and Imafidon (2009) highlighted the data security process and methods of the two most common security mechanisms available in WAPs. Tepsic, Veinović and Uljarević (2014) conducted an experiment on the performance of WPA2 in wireless networks. Whereas, Tsekleves and Tsekleves (2017) exposed the vulnerabilities of WPA2. These articles suggest that WPA2 protocol, though more secure than preceding protocols, have shortcomings in preventing unauthorized access to the network via its distributed keys. The exposé deliberately places WPA2 in WAPs as no longer adequate to offer wireless network security unless strengthened. Nevertheless, the Philippine government led no initiatives to enhance cybersecurity awareness for its people or at least tap the technology sectors in the country to assess vulnerabilities, to evaluate technical capability, and to design strong IT infrastructure to mitigate and counter these risks.

## *Phishing Attacks in the Philippines*

A phishing attack sends malicious emails (usually claiming to be from a bank) or fake websites that appear exactly the legitimate site devised in clever ways to fool clients into handling over valuable and sensitive information, such as usernames and passwords, and credit card information (Dodge, Carver, & Ferguson, 2007). Preventing these attacks can be tricky as these breaches invoke the human element (Thompson, 2013), such as carelessness, ignorance, fear, curiosity, urgency, or similar behaviors. These human factors play an ever-increasing role in the upsurge security attempts and cybercrime activities to a large degree (Von Solms & Van Niekerk, 2013).

In the Philippines , the bank industry (and its clients) seems to be the most favorite target, as fraud credit card transactions and other online scam cases reach more than half a billion pesos in 2016 (Jesus, 2017). Due to the mounting of fraud, two of the largest banks in the Philippines issued statements to warn their clients to be mindful over fake log-in sites, forged embedded forms, and suspicious banking procedures (ABS-CBN News, 2018a, 201b). The fake login site replicated the layout of thelogin page of a bank using a different website link. The event increased the number of claims for unauthorized withdrawals and other transactions. Security experts also advise against using public WiFi hotspots to prevent leakage of sensitive information. According to Microsoft Asia Pacific Security Intelligence Report (Microsoft, 2017), the Philippines ranked as the eighth most vulnerable in the Asia Pacific region to phishing or malware attacks with a 19.2% encounter rate as of March 2017.

## *Internet Security Awareness*

Developed countries promote cybersecurity awareness by way of shared endeavor. This is achieved through government, business industries, and academic partnerships. In the United States, cybersecurity is a large industry with a clear goal,that is, keep America safe (Department Homeland Security, 2017). They declared October as the National Cyber Security Awareness Month, which is also adopted by other countries such as Canada, Australia, and other European countries. The annual campaign is a collaborative effort between the government and industries to educate citizens and businesses with tools and resources to protect themselves online. Canada laid out a cybersecurity strategy based on five principles solely to maximize the benefits of Internet technology for Canadians. The Canadian government also has another proactive, collaborative strategy to cyber secure future — embed cybersecurity in the education system to foster a culture of security at the first exposure of Internet (Government of Canada, 2010).

It is clear that these countries are committed to protect their future from cyber attacks, as shown in their strategies, actions, and accomplishments. Following these relevant initiatives, South Africa (SA) also made cybersecurity efforts with the intent to secure cyber infrastructure and cultivate a culture of cybersecurity awareness amongst their people through South African Cyber-Security Academic Alliance (SACSAA). The goal of SACSAA is to engage primary schools by way of contest campaign awareness (Kortjan & Solms, 2014). The same is also underway in other countries such as Nigeria (Adelola, Dawson, & Batmaz, 2015), Malaysia (Hasan et al., 2015), and Saudi Arabia (Ramalingam, Shaik, & Mohammed, 2016). In contrast to South Africa, the Philippines is lagging behind in terms of tangible collaborative accomplishments to deal with cybersecurity awareness. According to Global Cybersecurity Index (International Telecommunication Union, 2017), the Philippines is low in the cybersecurity training pillar and must improve further in cooperation pillar. However, the Philippines gained a medium score in terms of cybercrime strategy and legislation (International Telecommunication Union, 2017). In 2004, the Philippines released a National Cyber Security Plan with the intention to provide a strategic and operational direction against cyber crime (Department of Information and Communications Technology (DICT), 2004). To effectively deal with cybersecurity, laws were passed in 2010 (RA8792 Electronic Commerce Act) and 2017 (RA10175 The Data Privacy Act), but the implementation is low and slow. Therefore, with the creation of the Department of Information and Communications Technology (DICT), mechanisms to develop a culture of security among Filipinos are much to be anticipated.

## METHODOLOGY

### *Data Source*

The Internet subscribers of Iriga Telephone (IrigaTel) Company Inc, Iriga City, Camarines Sur, Philippines were the respondents in the study. IrigaTel is a local telephone and Internet service provider in Camarines Sur, which delivers sophisticated broadband and landline services to consumers, and small and medium businesses. A total of 252 consented subscriber-respondents answered the questionnaire in its entirety from April 17 to May 23, 2017. The consented respondents were recruited based on non-probability quota sampling.

### *Research Approach*

Using one survey instrument, data were collected quantitatively combining several methods: traditional paper-based and online questionnaires, and phone interviews. The survey instrument is a two-page questionnaire comprised of refined questions related to the Internet usage and security practices on the Internet developed and distributed to the respondents. The questionnaire covered basic demographic information, Internet usage, and security practices. Internet usage element includes questions that identify how often users access the Internet, devices used to access the internet, the purpose of usage and place of access. Security practices element contains questions about the perceptions, habits, and attitudes of users regarding Internet security risks, protection and management. The internal consistency of the test items in the questionnaire was analyzed with Cronbach's Alpha of 0.9689 with adjectival rating of Excellent.

The accomplishment of the specific objectives required the use of frequency and percentage following Likert's five-point range and description to interpret the result. The 5-point scale, ranging from Strongly Agree (SA), Agree (A), Neutral (N), Disagree (D) and Strongly Disagree (SD), was used to allow respondents to express how much they agree or disgree on a given statement. Survey results have been processed and analyzed in several dimensions such as tables and graphs to thematically interpret them through narrative approach.

## RESULTS AND DISCUSSION

### *Internet Usage*

The percentile result of the demographic data in Figure 1 represents a fairly even split between male (44.4% of the total 252 respondents) and female (55.6% of the total respondents). It was found that 50% of the respondents have been on the Internet for 1-3 years while 27.8% and 16.7% of the respondents have been on the Internet for 4-6 years and over ten years, respectively. The sample reflects that the last three years has the fastest Internet growth rates. This confirms the report of Miniwatts Marketing Groups (2017) on the recent Internet growth statistics report in the Philippines.

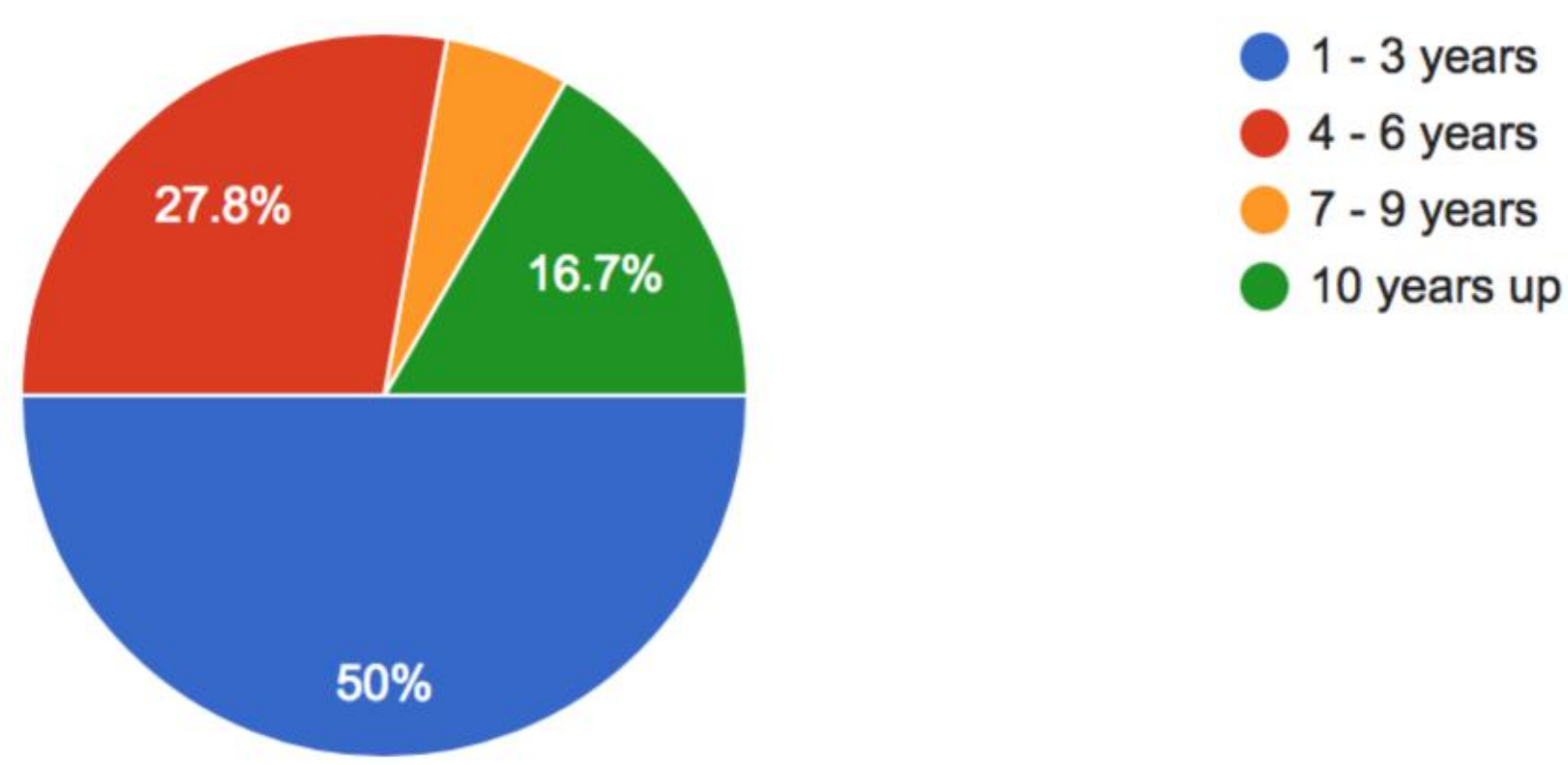

*Figure 1*. Years of Internet Usage

The study sought to establish the most popular device used to access Internet and places for Internet access, which may help form important strategic and operational decisions. As evidence, via Figure 2, 94.4% of respondents accounted for smartphones as the most preferred device to access the Internet, second in rank are laptop computers with 77.8% while smart tablets ranked third with 50%.

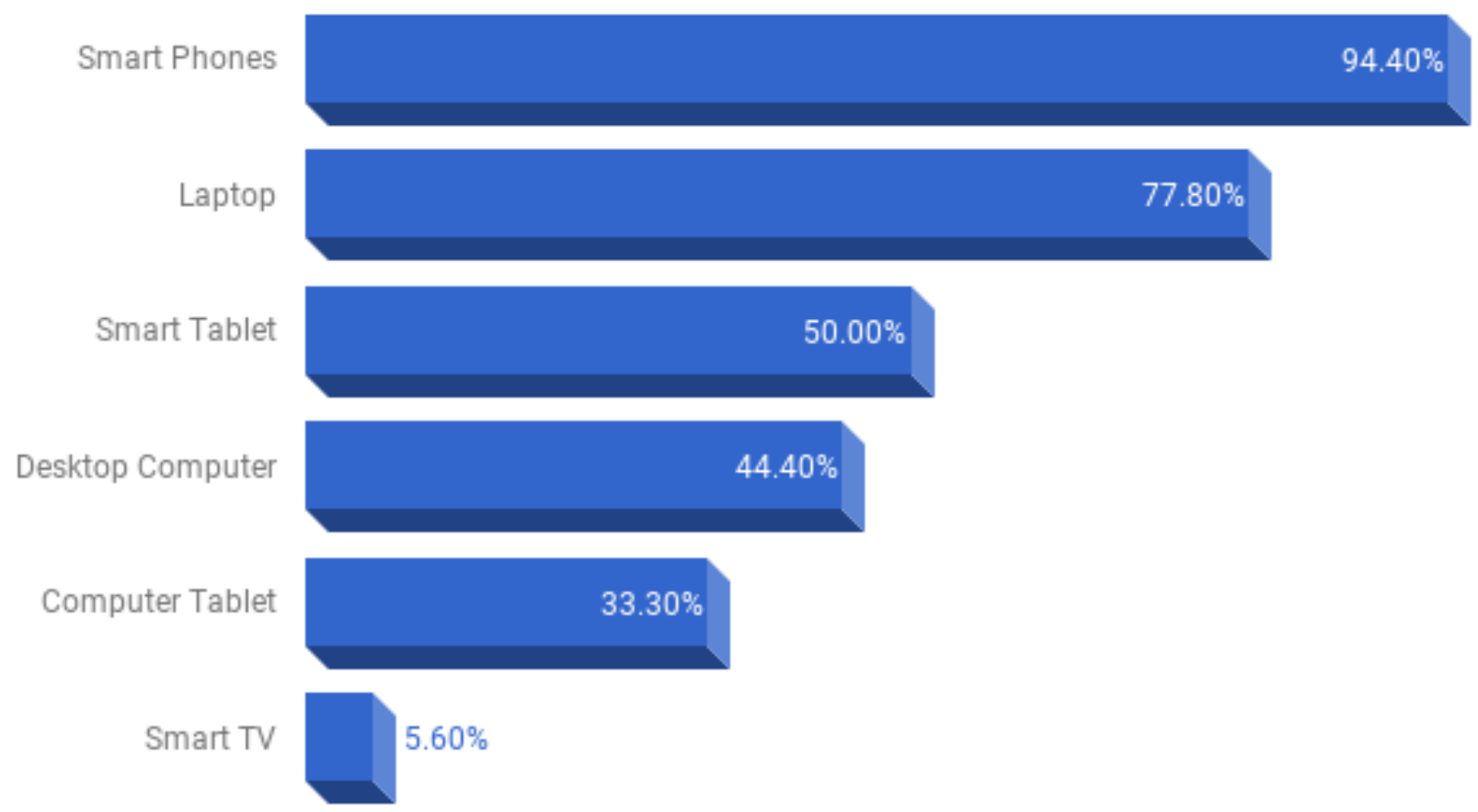

*Figure 2*. Leading devices used to access the Internet.

Survey findings also shed further light on the locations for Internet access that was not between fixed but rather mobile, as shown in Figure 3.

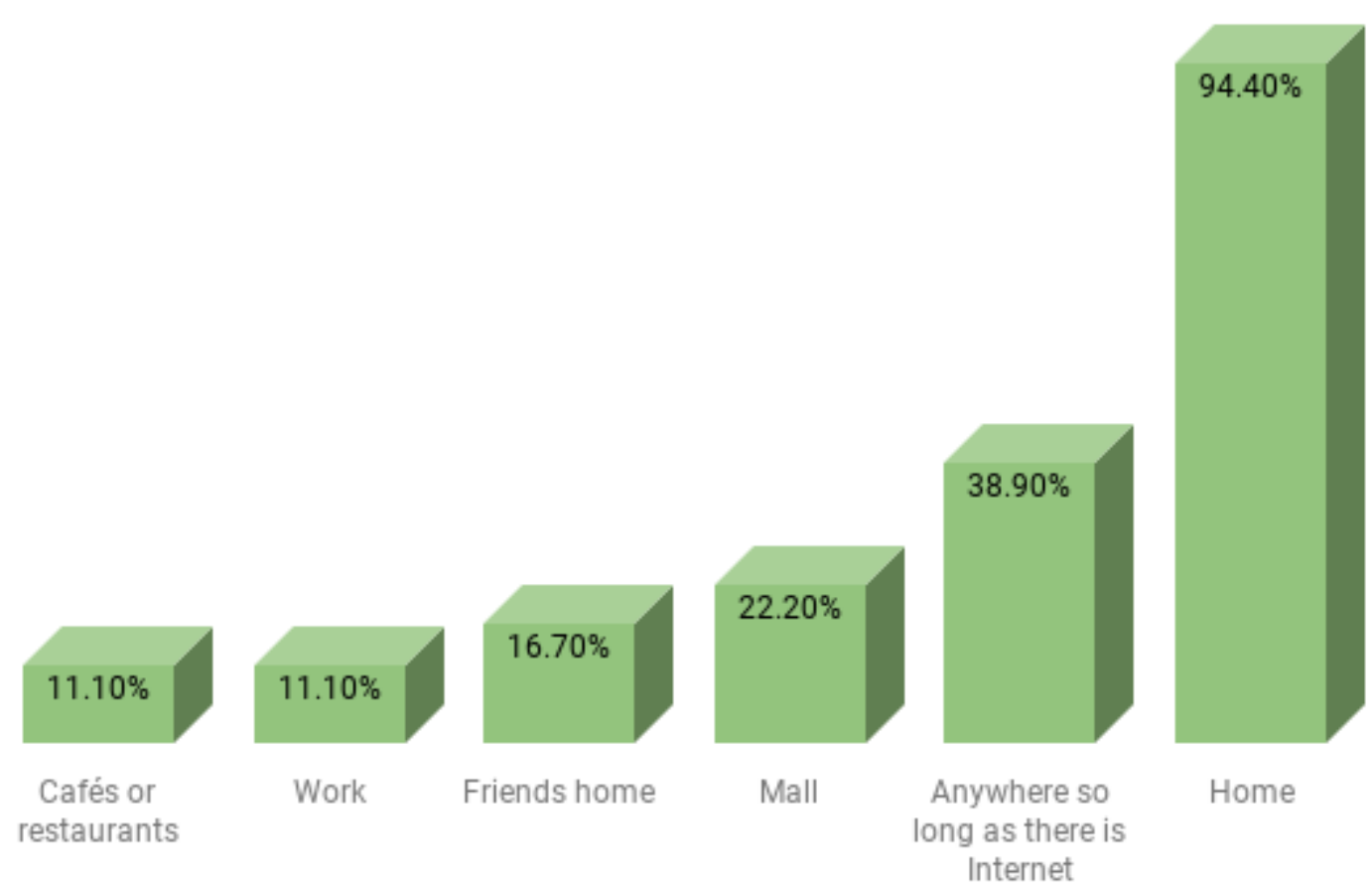

*Figure 3*. Places for Internet Access.

As far as places of access are concerned, at-home Internet access turned out to be the most frequent location for Internet access with 94.4%. However, it is also surprising to note the 38.9% and 22.2% Internet access are accessed "anywhere" and at the mall, respectively. These findings mean that Internet access was done outdoors. This implies that respondents access the Internet using public wireless connection regardless of the security or threats at hand.

This survey also focused on determining the activities respondents often use on the Internet. Rankings shown in Figure 4 reveal that the first ten (10) most popular internet activities by the respondents are as follows: (1) web browsing (88.98%), (2) email service (77.8%), (3) software download and use of social networking sites (72.2%), (4) chat with friends, shop online, and file sharing (66.7%), (5) send/ receive photos (61.7%), (6) play music (55.6%), (7) search/ apply job, read/ watch news and find people you know (44.4%), (8) play games and research book/ articles (38.9%), (9) bank online (33.3%), and (10) find local events (22.2%).

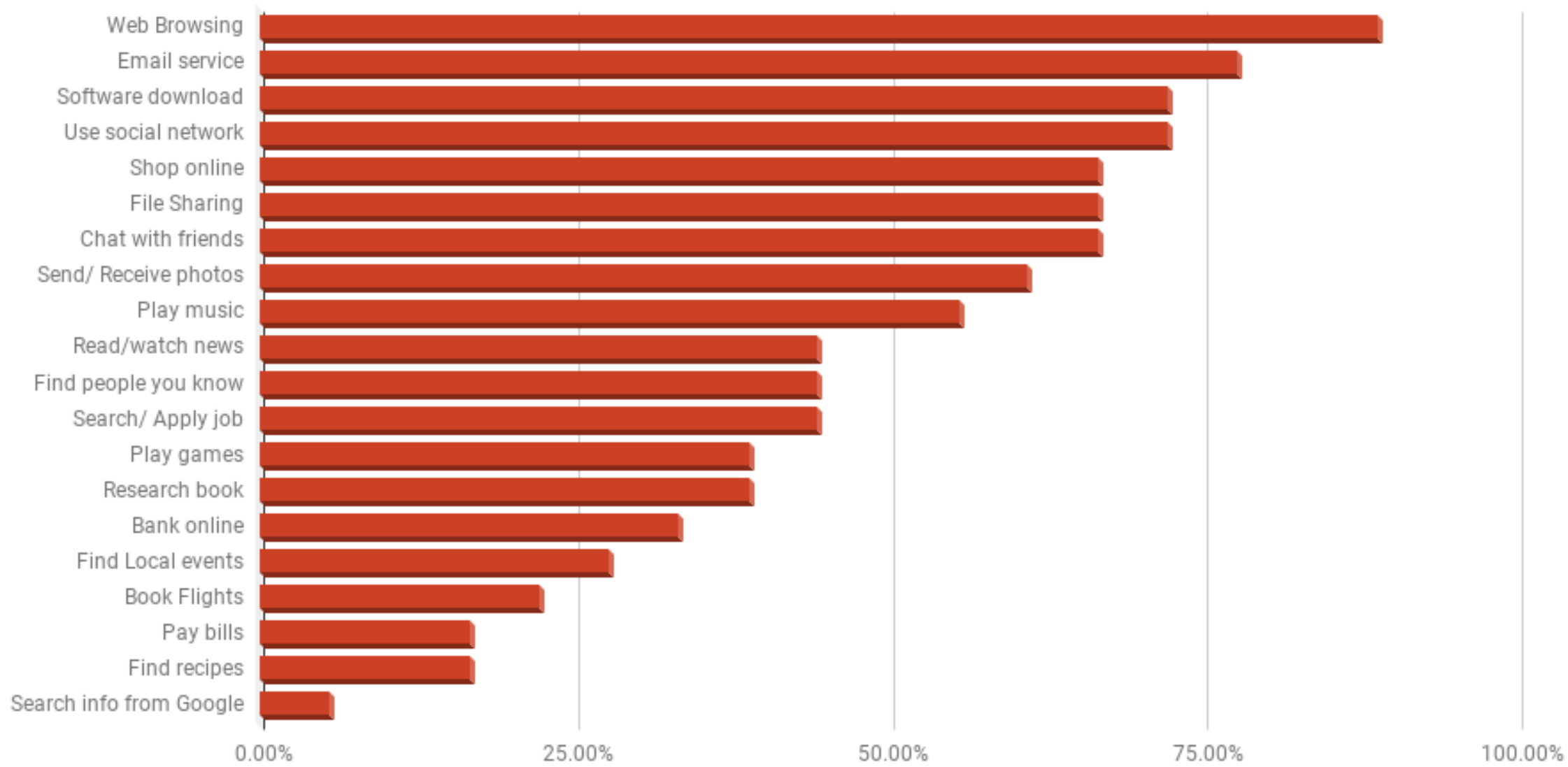

*Figure 4*. Percentage of activities often used on the Internet.

As described in the result, user motivation for Internet access suggested that Internet is used to expand general knowledge by browsing the free and diverse information available online, for communication, and widen social connectivity, which is in line with the result of Haridakis and Kim (2009). This, in turn, can lead to understanding what policy or programme the government can employ given the diverse form of usage. However, that perception is towards the optimistic secure and positive side because email service and software download can become a potential risk. A wide variety of malicious programs, attacks and social engineering tactics can easily be distributed using email service while downloading from a non-trustworthy source is the best direct way of transmitting and distributing malware, Trojans, and other viruses.

## *Security Practices*

In attempt to examine respondents' Internet security practices, the survey asked the question on the top three security protection measures they use. The results shown in Figure 5 indicates that the three most utilized security measures are (1) use of default security software of the computer and installed antivirus software (83.3%),(2) installation of a firewall (50%), and (3) installation of anti-spyware or anti-phishing software (33.3%).

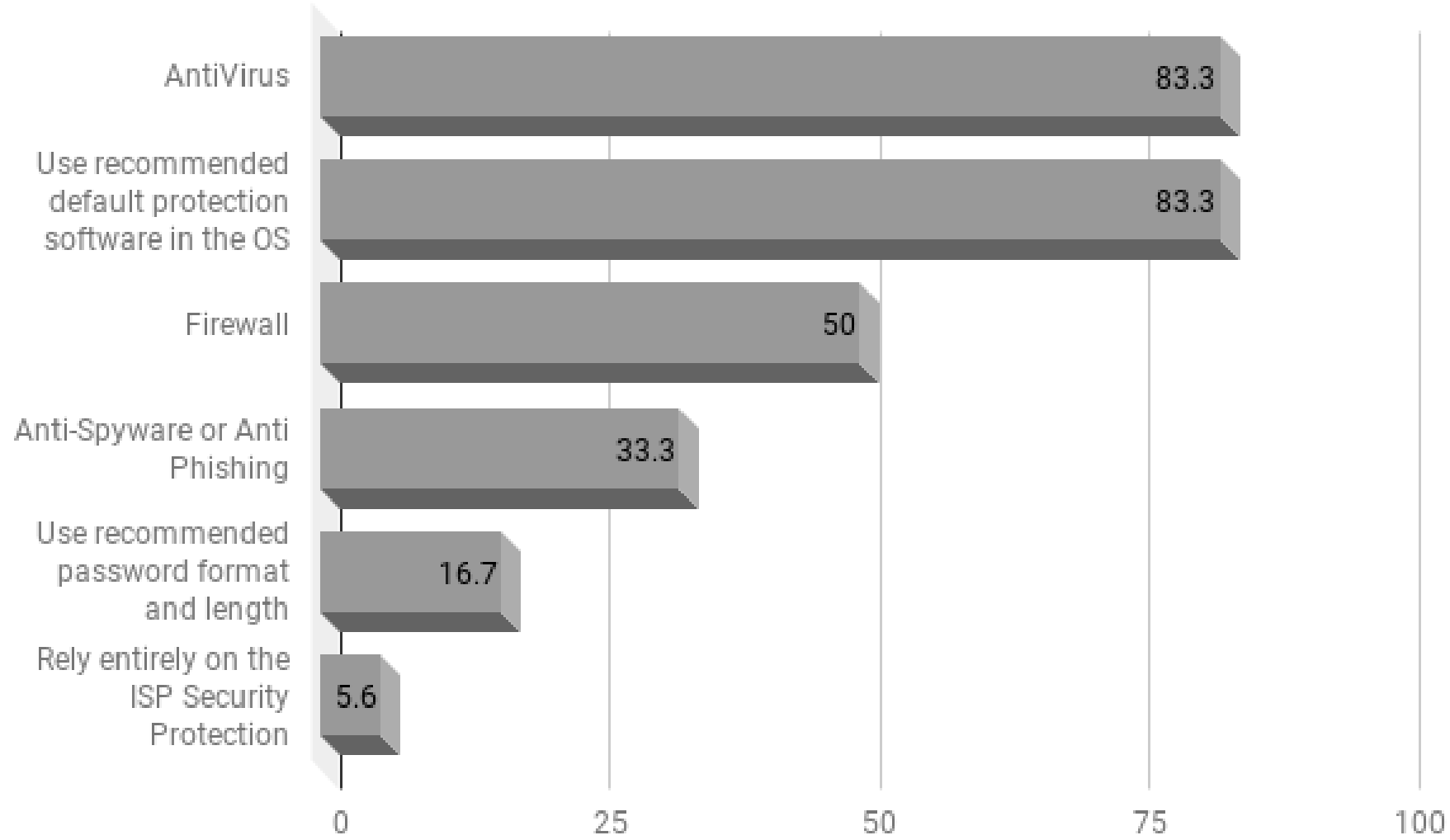

*Figure 5.* Percentage of Security Measures Used

The result in Figure 5 suggests that most respondents are wary about security software or sees the value of using the recommended password format. However, the result is also problematic with respect to the top three security software used. Operating systems usually include default antivirus and firewall. Installing another antivirus or firewall, then running them at the same time is not recommended. This may slow down or, worse, cause system conflict. It is also an unwise safety measure for respondents to just rely entirely on the Internet service provider security software. The default security software is the minimum standard; therefore, it comes with limited features (Howe, Ray, Roberts, Urbanska, & Byrne, 2012).

Table 1. Perception on Internet Risks

| Internet practices | SA | A | N | D | SD |
|---|---|---|---|---|---|
| Use of free WiFi may have security problems | 5.56% | 27.78% | 55.56% | 5.56% | 5.56% |
| Use electronic money or Internet banking anywhere | 5.56% | 27.78% | 38.89% | 27.68% | 0% |
| Leave the WiFi access point configuration unchanged | 27.78% | 50.00% | 5.56% | 5.56% | 11.11% |
| Use more than the recommended password length and format. | 27.78% | 50.00% | 11.11% | 0% | 11.11% |
| Security measures are up to the provider, not the individual. | 11.11% | 33.33% | 27.78% | 16.67% | 11.11% |

In respect to the statement "use of free WiFi may have security problems", Table 1 shows that more than half of the respondents (55.56%) were neutral to the statement. This level of response is somewhat disturbing. Thus, it could be inferred that most respondents are unaware of the risks using public WiFi. In case of "using electronic money or Internet banking anywhere", 38.89% (majority) responded neutral. On the other hand, half of the respondents (50%) agreed to the statement "use more than the recommended password length and format". The results suggest that a moderate proportion of respondents use online banking and purchasing over a public access point, which is unsafe. However, setting a strong password may be helpful. Meanwhile, majority of the respondents (50%) leave the configuration of the router unchanged, whereas about one third (33.3%) agreed that security measures are up to the provider. These suggest that respondents are neither concerned nor troubled about the possible security risk when default configuration in routers was left unchanged.

Finally, the respondents were asked to scale the growth of their Internet security awareness over the past years, majority (42.1%) answered that the awareness level did not change much or is still limited over the past years. This result implies that for the past years, majority have not had Internet security awareness training or very little self-enrichment is done by the respondents to increase their security awareness.

## CONCLUSIONS AND RECOMMENDATIONS

Driven by the genuine desire to bring public consciousness that Internet security is a serious matter, a survey was conducted to study the Internet usage and security awareness of Filipinos. Based on the findings, the number of Internet users has increased by 50% from 2014-2017, while smartphones continue to be the digital device of choice for Internet users. Respondents also do not mind using public access points available in outdoor locations such as in airports, restaurants, and parks, even when public networks are unsecured and dangerous. This finding underscores different implications, namely, (1) most mobile data plans include data cap and data coverage fees that are costly, (2) data mobile coverage is not everywhere, and/or (3) Internet security is only secondary over Internet connectivity. Also notable are the Internet activities that topped the rankings particularly email and software download. The risks of downloading files and opening emails from unknown users are tricky. Users might not find anything wrong in their computers until their computers systems show manifestations of virus infections. Internet banking also ranked 9th in the list of Internet activities, which was previously perceived negatively in society (Barquin & HV, 2015). This also suggests a new form of risk if not immediately negated through intervention. One interesting implication is that respondents do not have the intimate knowledge and comprehension of the Internet risks in their practical and day-to-day application.

In the case of security measures and Internet risks perception, the result showed several discrepancies on how respondents protect themselves online. This finding implies that, although respondents may have good knowledge of the Internet security software, proper implementation is very limited particularly when the security practice is acquired online or just taught by a colleague or friend, conveniently overlooking important facts. It may be true that one need not be an expert to protect oneself online but little knowledge is dangerous. This situation can be corrected through comprehensive and continuous awareness and training programs, and adoption of all the necessary safety measures that a non-expert must know.

Given the sharp increase of Internet users for the last three years and the very low awareness level identified among respondents, the Philippine government must act now and tap industries to educate Filipinos about Internet security before any negative consequences happen in the future. Without such education, Internet users will remain unmindful of the risks they face and unaware when to safeguard themselves. Meanwhile, ISPs may pioneer initiatives of bundling security defense tools with their services as the initial response to this concern. Users may raise their general level of Internet security awareness and self-protection through reading IT security magazines, books and online articles regularly. The academe should also consider including cybersecurity in the curriculum and extension services. Finally, in terms of technical perspective, security scientists and researchers must consider enhancing or developing a more reliant security mechanism to replace the protocols embedded in the standard WAPs as they are already weak and out of date.

## ACKNOWLEDGEMENT

My sincere appreciation and thanks to the Iriga Telephone Company Inc. management, especially to Atty. Butch Ortega and Engr. James Balase, CEO, and mentor, respectively, for their warm welcome and assistance throughout this research. Your support and acceptance enabled my study to be possible and representative.